\documentclass[conference]{IEEEtran}
\IEEEoverridecommandlockouts
\usepackage{cite}
\usepackage{amsmath,amssymb,amsfonts}
\usepackage{algorithmic}
\usepackage{graphicx}
\usepackage{textcomp}
\usepackage{xcolor}
\def\BibTeX{{\rm B\kern-.05em{\sc i\kern-.025em b}\kern-.08em
    T\kern-.1667em\lower.7ex\hbox{E}\kern-.125emX}}

\usepackage{booktabs}
\usepackage{tabularx}
\usepackage{longtable} 
\usepackage{colortbl}
\usepackage{multirow, multicol, array, makecell}
\usepackage{siunitx}
\usepackage{cleveref}
\usepackage{flushend}
\usepackage{subcaption}
\crefname{section}{Section}{Sections}
\crefname{figure}{Figure}{Figures}
\crefname{table}{Table}{Tables}

\usepackage{glossaries}
\newacronym{cnn}{CNN}{Convolutional Neural Network}
\newacronym{dl}{DL}{Deep Learning}
\newacronym{ai}{AI}{Artificial Intelligence}
\newacronym{dnn}{DNN}{Deep Neural Network}
\newacronym{nn}{NN}{Neural Network}
\newacronym{mmf}{MMF}{Multimedia Forensics}
\newacronym{vc}{VC}{Voice Conversion}
\newacronym{tts}{TTS}{Text-to-Speech}
\newacronym{ann}{ANN}{Artificial Neural Network}
\newacronym{df}{DF}{Deepfake}
\newacronym{ml}{ML}{Machine Learning}
\newacronym{roc}{ROC}{Receiver Operating Characteristic}
\newacronym{auc}{AUC}{Area Under the Curve}
\newacronym{mlp}{MLP}{Multi-Layer Perceptron}
\newacronym{ssl}{SSL}{Self-supervised Learning}
\newacronym{eer}{EER}{Equal Error Rate}
\newacronym{ba}{BA}{Balanced Accuracy}

\begin{document}

\title{
Synthetic Speech
Attribution \\ via Prototypical Networks

}

\author{
\IEEEauthorblockN{
Viola Negroni,
Paolo Bestagini,
Stefano Tubaro
}\vspace{0.35em}
\IEEEauthorblockA{\textit{Department of Electronics, Information and Bioengineering (DEIB), Politecnico di Milano}\\
\{viola.negroni, paolo.bestagini, stefano.tubaro\}@polimi.it}
}

\maketitle

\begin{abstract}
Synthetic speech attribution aims to identify the generative system responsible for a speech signal, but current approaches typically rely on black-box neural networks that provide limited insight into their decisions. 
This work investigates prototype-based networks as an interpretable alternative, where predictions are grounded in comparisons with representative training examples. 
We adapt ProtoPNet to spectrogram-based speech representations and evaluate the proposed framework on the MLAAD dataset under closed-set, cross-lingual, and open-set conditions. 
Experiments show that prototype-based reasoning achieves competitive or improved attribution performance compared with the baseline while enabling example-based explanations. 
These results highlight that interpretability and performance can be jointly achieved in synthetic speech attribution through prototype-based modeling.
\end{abstract}

\begin{IEEEkeywords}
synthetic speech, deepfake attribution, source tracing, interpretability, XAI, prototypes, TTS, audio forensics
\end{IEEEkeywords}

\section{Introduction}
\label{sec:intro}
Recent advances in text-to-speech (TTS) synthesis have made it possible to generate speech that is virtually indistinguishable from a real human voice.
While these systems enable numerous legitimate applications, they also raise serious concerns, as \textit{deepfake} speech can be misused for impersonation, fraud, and disinformation~\cite{amerini2026deepfake}.
Considerable research effort has therefore focused on speech deepfake detection, i.e., determining whether a given signal is real or synthetically generated~\cite{borrelli2021synthetic, wang2022investigating, salvi2024comparative, negroni2024leveraging, ge2025post}.
However, a binary real-versus-fake decision is often insufficient in forensic and investigative contexts, where identifying \emph{which} system generated a signal provides considerably more actionable information.
This task is referred to as synthetic speech attribution or source tracing.

Existing approaches to source tracing typically rely on deep neural classifiers trained to discriminate among a set of known generators~\cite{salvi2022exploring, zhu2022source, deng2024vfd, klein2024source, stan2025tada}.
While effective, these models are black-box in nature. 
Their predictions offer no direct insight into the acoustic evidence supporting a given attribution.
Closer to our setting, Mishra et al.~\cite{mishra2026towards} propose an explainable framework for source tracing based on probabilistic attribute embeddings, using Shapley values to quantify each attribute's contribution to the decision.
Although this provides interpretable input representations, the underlying classifier remains black-box, and explanations are produced post-hoc rather than being a structural property of the decision process itself.
As argued in~\cite{rudin2019stop}, post-hoc explainability is intrinsically flawed, as explanations produced after training are not guaranteed to reflect the model's true decision process, and can therefore be misleading rather than clarifying.

\begin{figure}[t]
    \centering
    \includegraphics[width=\columnwidth]{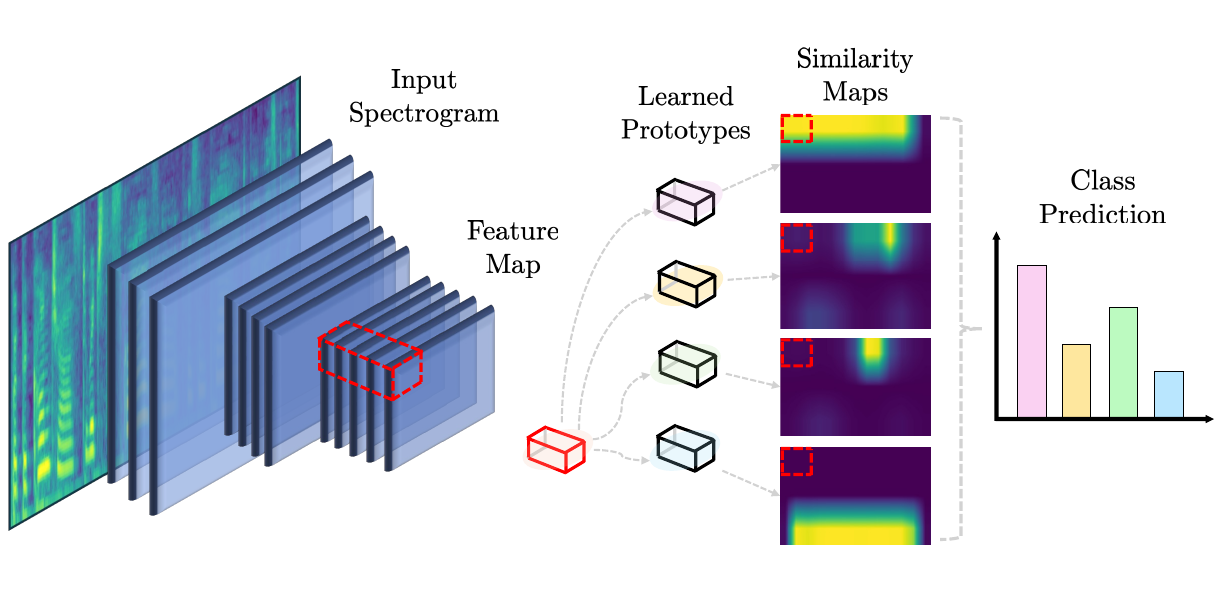}
    \caption{Overview of the proposed prototype-based framework for interpretable synthetic speech attribution.}
    \label{fig:overview}
\end{figure}

In this work, we investigate an interpretable-by-design approach to synthetic speech attribution based on prototypical networks~\cite{chen2019looks}, in which explanations are the direct output of the classification mechanism, rather than an external approximation of it (\Cref{fig:overview}).
Specifically, we ask: (i) whether prototype-based networks can match the performance of standard black-box classifiers on synthetic speech attribution; (ii) if alternative ways of grouping generators can provide a more effective and more generalizable formulation of the task; and (iii) how prototype-based reasoning behaves under cross-lingual and open-set conditions, where training and evaluation data diverge from the closed, single-language setting.

Our contributions are as follows.
We adapt ProtoPNet~\cite{chen2019looks}, originally proposed for image classification, to spectrogram-based synthetic speech attribution.
We evaluate the proposed framework on the MLAAD dataset under closed-set, cross-lingual, and open-set scenarios, and provide a quantitative analysis of the learned prototypes, showing that their acoustic focus and usage patterns are interpretable and consistent with domain knowledge on synthetic speech artifacts.

\section{Proposed Method}
\label{sec:method}
This section introduces the proposed framework for source tracing, including the task formulation, model architecture, training objective, and prototype grounding procedure.

\subsection{Problem Formulation}
\label{subsec:problem}
The synthetic speech attribution problem, also referred to as source tracing, is formulated as a multi-class classification task. 
Given a discrete-time synthetic speech signal $\mathbf{x}$ and a set of $N$ synthesis methods $Y=\{y_0,\dots,y_{N-1}\}$, the objective is to predict the generator $y_i \in Y$ used to synthesize $\mathbf{x}$.

In the closed-set scenario, all $N$ classes are observed during both training and evaluation. 
In the open-set scenario, a subset of synthesis methods is absent from training and introduced only during evaluation. 
The model must therefore detect and reject samples from unknown classes.

\subsection{Proposed System}
\label{subsec:system}

\subsubsection{Interpretability by Design}
\label{subsubsec:system_1}
Most explainability methods for deep neural networks are \emph{post-hoc}.
They approximate or probe an already trained black-box model to produce an explanation afterward (e.g., GradCAM~\cite{selvaraju2017grad}).
Such explanations describe \emph{where} a model looked, but not necessarily \emph{why} that evidence was considered relevant~\cite{rudin2019stop}.
The underlying decision process remains unclear.
Prototype-based networks follow a different paradigm, sometimes referred to as \emph{ante-hoc} or \emph{interpretable-by-design}.
Rather than explaining a black box, the model itself is constrained, at training time, to reason through explicit comparisons to a set of learned reference patterns, or \emph{prototypes}.
At inference time, a prediction is the direct result of these comparisons, and can therefore be reported as ``this input looks like that stored example, particularly in this region'', a form of case-based reasoning.
Interpretability is thus a structural property of the model, not an external approximation of it.

\subsubsection{Architecture Overview}
\label{subsubsec:system_2}
We adopt ProtoPNet~\cite{chen2019looks} for our experiments.
The system consists of two components: a \emph{backbone}, which extracts a feature representation from the input, and a \emph{prototypical layer}, which performs the case-based comparison described above.
The backbone is a standard, replaceable convolutional feature extractor.
Any architecture that produces a spatial feature map from a 2D input can be used in its place, and its specific choice is orthogonal to the interpretability mechanism itself.
The prototypical layer is the component that is added on top of it, and is what makes the overall system explainable.

Given an input $x$, the backbone $f(\cdot)$ produces a feature map
\begin{equation}
z = f(x), \quad z \in \mathbb{R}^{D \times H \times W},
\label{eq:backbone}
\end{equation}
where $D$ is the number of feature channels and $H \times W$ is the spatial resolution of the resulting map.

The prototypical layer holds a bank of $P = N_c \times K$ learnable prototype vectors $p_i \in \mathbb{R}^{D}$, $i = 1, \dots, P$, where $N_c$ is the number of training classes and $K$ the number of prototypes per class.
Each prototype is assigned to exactly one training class, encoded by a fixed assignment matrix.
For every prototype $p_i$ and every spatial location $(h,w)$ of $z$, we compute the squared Euclidean distance
\begin{equation}
d_i(h,w) = \lVert z(h,w) - p_i \rVert_2^2,
\label{eq:distance}
\end{equation}
implemented efficiently as a $1{\times}1$ convolution, and convert it into a similarity score via a monotonically decreasing, bounded transform
\begin{equation}
s_i(h,w) = \log\!\left(\frac{d_i(h,w) + 1}{d_i(h,w) + \epsilon}\right),
\label{eq:similarity}
\end{equation}
where $\epsilon>0$ is a small constant that prevents the logarithm from diverging as in~\cite{chen2019looks}.
The resulting similarity map $s_i \in \mathbb{R}^{H \times W}$ retains spatial information and can be used, upsampled to the input resolution, to visualize \emph{where} the prototype $p_i$ found its best match.
Global max pooling over the spatial dimensions yields a single activation value per prototype,
\begin{equation}
g_i = \max_{(h,w)} \, s_i,
\label{eq:pooling}
\end{equation}
and the resulting vector $g = [g_1, \dots, g_P]$ is passed to a final linear layer to produce the class logits.

\subsubsection{Training Objective}
\label{subsubsec:system_3}
As in~\cite{chen2019looks}, the network is trained by minimizing a composite objective defined as
\begin{equation}
\mathcal{L} =
\mathcal{L}_{\mathrm{CE}}
+ \lambda_{\mathrm{clst}} \mathcal{L}_{\mathrm{clst}}
+ \lambda_{\mathrm{sep}} \mathcal{L}_{\mathrm{sep}}
+ \lambda_{\mathrm{L1}} \mathcal{L}_{\mathrm{L1}},
\label{eq:total_loss}
\end{equation}
where $\lambda_{\mathrm{clst}}$, $\lambda_{\mathrm{sep}}$, and $\lambda_{\mathrm{L1}}$ are scalar weights balancing the contribution of each term, whose values are reported in \Cref{subsec:config}.
$\mathcal{L}_{\mathrm{CE}}$ is the standard cross-entropy loss on the predicted class.
The clustering loss
\begin{equation}
\mathcal{L}_{\mathrm{clst}} = - \max_{i \in \mathcal{P}_y} g_i
\label{eq:cluster_loss}
\end{equation}
encourages each sample to strongly activate at least one prototype assigned to its ground-truth class $y$, where $\mathcal{P}_y$ denotes the subset of prototypes belonging to class $y$.
The separation loss
\begin{equation}
\mathcal{L}_{\mathrm{sep}} = \max_{i \notin \mathcal{P}_y} g_i
\label{eq:separation_loss}
\end{equation}
penalizes high activation of prototypes belonging to other classes.
The regularization term $\mathcal{L}_{L1}$ applies an $\ell_1$ penalty to the
weights of the final classification layer, promoting sparse connections between
each class and prototypes of other classes, so that predictions rely primarily
on positive, own-class evidence.

\subsubsection{Prototype Grounding}
\label{subsubsec:system_4}
Training proceeds in three stages.
We first warm up the backbone and prototype vectors with the classifier frozen at a fixed initialization, then jointly train all parameters.
Periodically, we \emph{push} each prototype onto the closest same-class patch found in the training set, replacing its (possibly abstract) learned vector with the exact feature vector of a real training example.
This step grounds each prototype in an inspectable, real audio sample, rather than an arbitrary point in feature space, and is what allows predictions to be explained via direct comparison to a specific training utterance (\Cref{subsec:res_2}).
After each push, we briefly fine-tune only the final classification layer to recalibrate it to the updated prototypes.
\section{Experimental Setup}
\label{sec:setup}
This section describes the dataset, model configuration, training procedure, and evaluation metrics used in our experiments.

\subsection{Dataset and Partitioning}
\label{subsec:dataset}
MLAAD~\cite{muller2024mlaad} is a large-scale multilingual dataset comprising exclusively synthetic speech signals. 
The corpus is continuously updated to include samples generated by newly emerging \gls{tts} systems. 
In this work, we use MLAAD v8.
Since MLAAD does not provide predefined data partitions, we define our own splits for the attribution task. 
We train our source tracing model using only the English subset of MLAAD. 
This choice aims at reflecting a realistic scenario, as English-based corpora are predominant in the development and benchmarking of speech technologies. 
The remaining multilingual data are reserved for cross-lingual evaluation, which is discussed separately in \Cref{subsec:res_3}.
The English subset of MLAAD v8 contains \num{68} \gls{tts} model instances, i.e., synthetic speech generators, with \num{1000} tracks per class. 
To evaluate open-set generalization, we reserve a subset of these classes for evaluation. 
Specifically, we select the classes introduced between MLAAD v7 and v8, as they correspond to previously unseen \gls{tts} systems and simulate the emergence of new generators in real-world deployments.
The remaining \num{54} classes are used to construct the training, validation, and closed-set evaluation splits. 
For each class, \num{70}\% of the tracks are allocated for training. 
The remaining \num{30}\% are further divided, with \num{60}\% used for validation and \num{40}\% reserved for closed-set evaluation. Splits are class-stratified.

In addition to the standard grouping by \gls{tts} model instances, we investigate an alternative grouping strategy based on \gls{tts} families (see \Cref{subsec:res_1}). 
This analysis aims to determine whether grouping related generators provides a more effective formulation of the source tracing task. 
To this end, based on the available metadata, we merge generators that differ only in their training data, as well as generators that belong to the same architectural family but differ in model size. 
The same training, validation, closed-set, and open-set partitioning procedure described above is then applied to the resulting classes. 
With this grouping strategy, the number of training classes is reduced from \num{54} to \num{44}, while the number of open-set classes decreases from \num{14} to \num{11}.

\subsection{Model Configuration and Training Details}
\label{subsec:config}
During training, input segments are extracted as random \SI{4}{\second} windows from each utterance. 
If an utterance is shorter than \SI{4}{\second}, the signal is repeated to reach the target duration. 
We peak-normalize samples and trim leading and trailing silences to avoid potential shortcut learning~\cite{muller21_asvspoof}.
Potential class imbalance is addressed through balanced sampling. 
For prototype pushing, validation, and evaluation, we use the central \SI{4}{\second} window of each utterance to ensure determinism.
All audio signals are resampled to \SI{16}{\kHz}. 

We employ a ResNet18~\cite{he2016deep} backbone on top of the prototypical layer, and refer to the resulting model as ProtoRN18.
Originally designed for image classification, ResNet18 has become a widely used baseline for synthetic speech detection on 2D spectrogram representations, and was recently established as a benchmark architecture for synthetic speech attribution~\cite{klein2024source, negroni2025source}. 
The input consists of an 80-bin mel-spectrogram computed with a \num{512}-point FFT, using a \SI{25}{\milli\second} window and a \SI{10}{\milli\second} hop length, resulting in feature maps of size $80\times401$. 
The backbone follows the standard ResNet18 architecture, with an initial convolutional block followed by \num{4} groups of residual layers. 
The first convolutional layer is adapted for single-channel inputs by averaging the pretrained RGB filters across channels.
The feature map resolution is progressively reduced through the residual layers, reaching $5 \times 26$ after the third group and $3 \times 13$ after the fourth.
We consider both as candidate truncation points, since the choice trades spatial
granularity of the resulting explanations against representational depth.
The final feature map is projected to $D = 128$ channels by an additional $1 \times 1$ convolutional layer (from $512$ channels when truncating after the fourth group, $256$ after the third), producing the embeddings used by the prototypical layer.
We denote configurations as {L}$k${P}$n$, where $k \in \{3, 4\}$ is the residual group after which the backbone is truncated, and $n \in \{1, 3, 6\}$ is the number of prototypes per class $K$ (e.g., L3P6 means group layer 3 with 6 prototypes).

The prototypical network is trained for \num{5} warm-up followed by \num{100} joint-training epochs using a learning rate of $10^{-3}$ with cosine annealing. 
Every \num{5} epochs, a prototype push operation followed by classification-layer fine-tuning is performed, to anchor the learned prototypes to actual training (\Cref{subsubsec:system_4}). 
The checkpoint achieving the highest validation accuracy is selected, after which a final prototype push and classification-layer fine-tuning step is applied.
The loss weights are set to $\lambda_{\mathrm{clst}}=0.8$, $\lambda_{\mathrm{sep}}=0.4$, and $\lambda_{\mathrm{L1}}=10^{-4}$.

\subsection{Evaluation Metrics}
\label{subsec:metrics}
\gls{ba} is used as the primary metric for closed-set evaluation.
We additionally report the macro F1 score, which provides a balanced measure of precision and recall across classes. 
For open-set evaluation, we report the \gls{auc} and FPR@TPR95, which measures the false positive rate when \num{95}\% of positive samples are correctly detected.
\section{Results}
\label{sec:results}
In this section we evaluate the proposed framework. 
First, we assess closed-set attribution performance and analyze the learned prototypes. 
We then investigate robustness under cross-lingual and open-set conditions.

\subsection{Closed-set Performance}
\label{subsec:res_1}

\begin{table}
\caption{Comparison between ResNet18 and ProtoRN18 on the model and architecture classification tasks.}
\label{tab:model_arch_results}
\centering
\resizebox{0.95\columnwidth}{!}{
\begin{tabular}{llcc}
\toprule
\textbf{Training by} & \textbf{Model} & \textbf{BA $\uparrow$} & \textbf{Macro F1 $\uparrow$} \\
\midrule
\multirow{3}{*}{TTS Instances}
& ResNet18          & 86.25\% & 0.8504 \\
& ProtoRN18 L4P3    & 87.48\% & 0.8674 \\
& ProtoRN18 L4P6    & 88.06\% & 0.8801 \\
\midrule
\multirow{3}{*}{TTS Families}
& ResNet18          & 99.06\% & 0.9905 \\
& ProtoRN18 L4P3    & 99.27\% & 0.9928 \\
& ProtoRN18 L4P6    & 99.65\% & 0.9964 \\
\bottomrule
\end{tabular}}
\end{table}

\begin{table}
\caption{Ablation study on ProtoRN18 configurations.}
\label{tab:protrn_ablation}
\centering
\resizebox{0.7\columnwidth}{!}{
\begin{tabular}{lcc}
\hline
\toprule
& \textbf{BA $\uparrow$} & \textbf{Macro F1 $\uparrow$} \\
\midrule
ProtoRN18 L3P1 & 98.93\% & 0.9893 \\
ProtoRN18 L3P3 & 98.97\% & 0.9893 \\
ProtoRN18 L3P6 & 98.83\% & 0.9902 \\
ProtoRN18 L4P1 & 98.86\% & 0.9887 \\
\midrule
ProtoRN18 L4P3 & 99.27\% & 0.9928 \\
ProtoRN18 L4P6 & 99.65\% & 0.9964 \\
\bottomrule
\end{tabular}}
\end{table}

In this experiment, we compare our framework against a ResNet18 baseline, consisting of the same backbone architecture without the prototypical layer, and trained with cross-entropy loss.
We evaluate both models on a held-out set containing samples from the same classes as the training data (closed-set evaluation).
We first compare two labeling granularities: grouping samples by the specific TTS instance (i.e., model checkpoint/version) versus by the underlying synthesis architecture family (see \Cref{subsec:dataset}).
As shown in \Cref{tab:model_arch_results}, architecture-level grouping yields a substantially higher \gls{ba} (e.g., 99.06\% vs.\ 86.25\% for the ResNet18 baseline).
This suggests that many instance-level distinctions are not reliably separable from the acoustic signal alone, while architecture-level distinctions are.
We adopt architecture-level grouping for the remainder of this work.

Notably, two of the prototypical network configurations, L4P3 and L4P6, exceed the ResNet18 baseline under this setting, an encouraging result given the added interpretability constraints on the representation.

\Cref{tab:protrn_ablation} reports an ablation over backbone truncation depth and number of prototypes per class.
While only L4P3 and L4P6 exceed the ResNet18 baseline (\Cref{tab:model_arch_results}), all other configurations still achieve strong performance, above \num{98.83}\% \gls{ba}.
Since neither L4P3 nor L4P6 dominates the other, we will be comparing both in the following.

\subsection{Prototypes Analysis}
\label{subsec:res_2}

\begin{table}
\caption{Distribution of learned prototypes activation \\ center across frequency regions.}
\label{tab:proto_distribution}
\centering
\resizebox{0.85\columnwidth}{!}{
\begin{tabular}{lcc}
\toprule
\textbf{Mel region} & \textbf{Prototype center} & \textbf{\# Source classes} \\
\midrule
Low-frequency  & $<1000$ Hz       & 1 (2.27\%) \\
Mid-frequency  & 1000--3000 Hz    & 5 (11.36\%) \\
High-frequency & $>3000$ Hz       & 38 (86.36\%) \\
\bottomrule
\end{tabular}}
\end{table}

\begin{figure*}[t]
    \centering
    \begin{subfigure}[b]{0.32\textwidth}
        \centering
        \includegraphics[width=\textwidth]{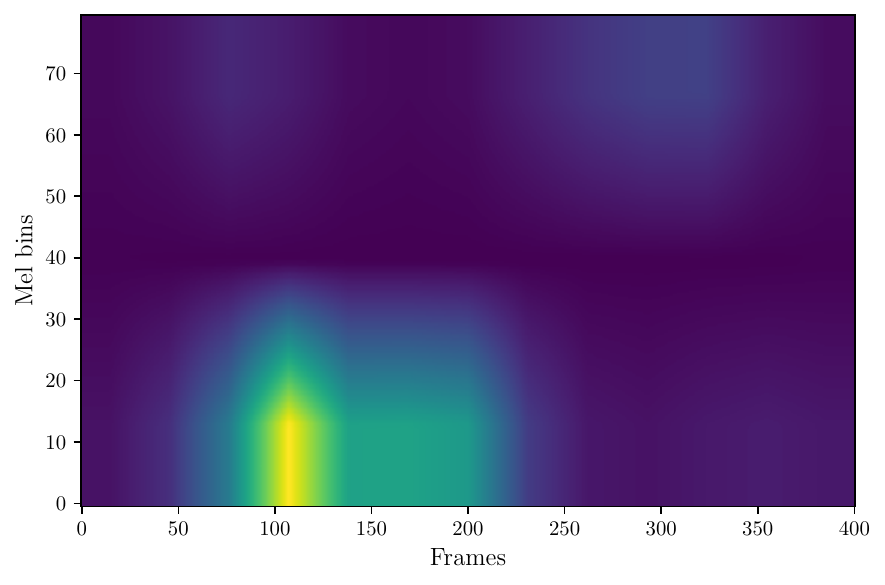}
        \caption{Prototype similarity map.}
        \label{fig:prototype_similarity}
    \end{subfigure}
    \hfill
    \begin{subfigure}[b]{0.32\textwidth}
        \centering
        \includegraphics[width=\textwidth]{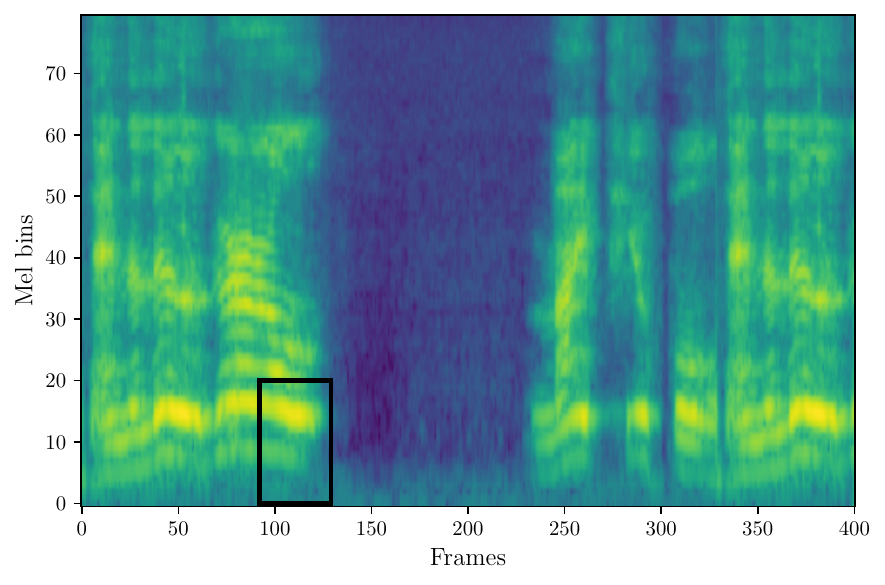}
        \caption{Match localization on input spectrogram.}
        \label{fig:prototype_localization}
    \end{subfigure}
    \hfill
    \begin{subfigure}[b]{0.32\textwidth}
        \centering
        \includegraphics[width=\textwidth]{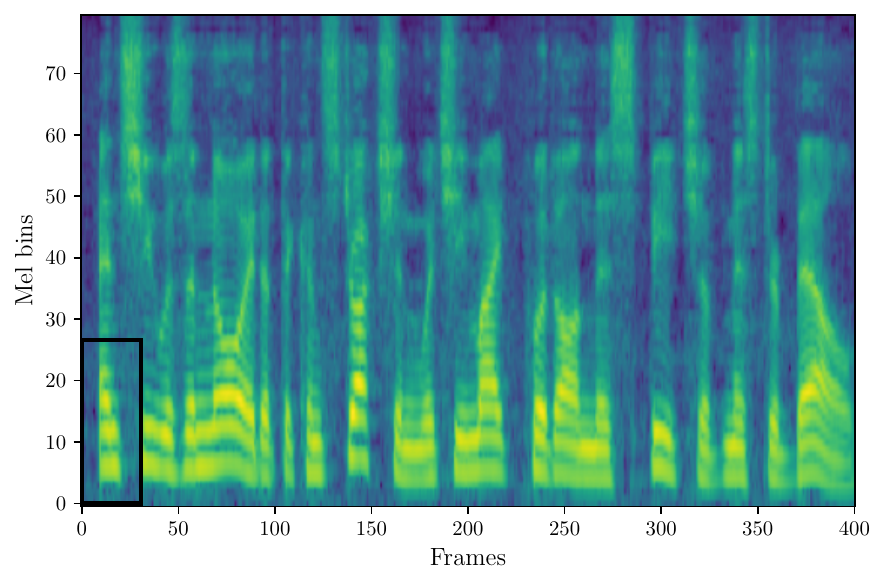}
        \caption{Grounded prototype spectrogram.}
        \label{fig:prototype_grounding}
    \end{subfigure}

    \caption{
        Prototype-based explanation for the test sample \texttt{jane\_eyre\_11\_f000168.wav} generated by \texttt{sesame\_csm}. The first panel shows the spatial similarity map between the input spectrogram and the selected prototype (higher values, stronger activation). The second localizes the maximum activation on the input spectrogram. The third shows the training spectrogram from which the prototype was grounded (\texttt{northandsouth\_40\_f000076.wav}) and its corresponding location.}
    \label{fig:prototype_explanation}
\end{figure*}

Push operations anchor the learned prototypes to existing regions of training samples (see \Cref{subsubsec:system_4}).
This enables direct inspection of the acoustic patterns represented by each prototype. 
By retrieving the source sample and activation location associated with each pushed prototype, we can quantitatively analyze where different classes are characterized in the spectrogram.
See the example in \Cref{fig:prototype_explanation}.

As shown in \Cref{tab:proto_distribution}, for the vast majority of source classes (\num{86.36}\%), the dominant prototype has an activation centroid located in the high-frequency mel region (above 3000 Hz). 
This observation is consistent with prior work showing that synthesis artifacts are predominantly concentrated in higher-frequency bands~\cite{salvi2023towards}. 
The speech synthesis process is commonly divided into two stages: acoustic modeling and vocoding. 
Such artifacts have been linked to neural vocoding~\cite{cuccovillo2026neural} and have been shown to provide stronger cues for source attribution than artifacts introduced by acoustic modeling~\cite{klein2024source}.
Only a small number of classes depart from this pattern, with \texttt{sesame\_csm} being the only class centered in the low-frequency region.

We note two caveats relevant to interpreting this result. 
First, the coarse three-region division of the frequency axis follows directly from the backbone's spatial output resolution at the L4 truncation point (see \Cref{subsec:config}).
In other words, the resulting explanation localization is bounded by the feature map resolution at which prototypes are derived.
Second, inspecting prototype usage per class revealed that, for most classes, a single prototype typically accounts for the large majority of top-contribution assignments across test samples.
This points to a degree of redundancy in the learned prototype set. 
The ablation in \Cref{subsec:res_1} shows that additional prototypes per class improve closed-set performance, so the remaining prototypes do have an effect on the classification boundary; however, they less frequently constitute the single most salient piece of evidence for a prediction.

\subsection{Cross-lingual Analysis}
\label{subsec:res_3}

\begin{table}
\caption{Cross-lingual performance.}
\label{tab:cross_lingual}
\resizebox{\columnwidth}{!}{
\begin{tabular}{lcccc}
\toprule
\textbf{Eval Setting} & \textbf{Model} & \textbf{Avg BA $\uparrow$} & \textbf{Avg BA Eng $\uparrow$} & \textbf{Drop $\downarrow$} \\
\midrule
\multirow{2}{*}{All classes}
& ProtoRN18 L4P3 & 60.78\% & 99.46\% & 38.68 pp \\
& ProtoRN18 L4P6 & 60.50\% & \textbf{99.71\%} & 39.21 pp \\
\midrule
\multirow{2}{*}{Filtered}
& ProtoRN18 L4P3 & 88.63\% & 99.53\% & 10.89 pp \\
& ProtoRN18 L4P6 & \textbf{89.07\%} & 99.63\% & \textbf{10.56 pp} \\
\bottomrule
\end{tabular}}
\end{table}

\begin{table}
\caption{ProtoRN18 L4P6 per-language performance.}
\label{tab:_frequent_lang}
\resizebox{\columnwidth}{!}{
\begin{tabular}{lcccc}
\hline
\toprule
\textbf{Language} & \textbf{TTS Families} & 
\textbf{BA $\uparrow$} & 
\textbf{BA Eng $\uparrow$} &
\textbf{Drop $\downarrow$} \\
\midrule
Portuguese & 5  & 96.66\% & 99.67\% & 3.01 pp \\
Spanish    & 9  & 89.00\% & 99.52\% & 10.52 pp \\
Italian    & 10 & 88.72\% & 99.48\% & 10.76 pp \\
Polish     & 7  & 87.98\% & 99.64\% & 11.66 pp \\
French     & 15 & 85.79\% & 99.55\% & 13.76 pp \\
Chinese    & 8  & 82.76\% & 99.45\% & 16.69 pp \\
German     & 10 & 80.62\% & 99.58\% & 18.96 pp \\
Dutch      & 6  & 78.72\% & 99.74\% & 21.02 pp \\
Japanese   & 5  & 73.70\% & 99.51\% & 25.81 pp \\
Russian    & 7  & 56.30\% & 99.54\% & 43.24 pp \\
\bottomrule
\end{tabular}}
\end{table}

Most speech deepfake datasets are primarily composed of English speech, but modern \gls{tts} systems can synthesize high-quality speech in multiple languages.
This makes cross-lingual generalization a relevant challenge.
Previous studies have shown that language mismatch between training and evaluation can degrade performance~\cite{negroni2025source, xuan2025multilingual}.
We therefore investigate this effect in our setting.

\Cref{tab:cross_lingual} reports the average per-language \gls{ba} over the \num{38} non-English MLAAD languages for ProtoRN18 L4P3 and ProtoRN18 L4P6.
In the ``All classes'' setting, both models show a substantial drop relative to the \textit{matched} English split, i.e., the subset of the English test set containing only generator classes also present in the target language.
We analyzed errors by language and generator family to identify the source of this degradation.
A substantial portion of the drop was associated with a small subset of classes whose training representation was limited to single-speaker \gls{tts} checkpoints derived from the LJSpeech dataset~\cite{LJSpeech}, namely \texttt{Glow-TTS}, \texttt{Tacotron2-DCA}, \texttt{Tacotron2-DDC}, and \texttt{VITS}.

To assess their impact, we repeated the evaluation after excluding these classes.
This removed four languages entirely (Slovenian, Croatian, Lithuanian, and Latvian), which no longer retained any target classes, reducing the evaluation set from \num{38} to \num{34} languages.
The performance drop now is substantially lower (\Cref{tab:cross_lingual}, ``Filtered'' setting).
This suggests that a large part of the observed language mismatch stems from limited speaker coverage in the training data, rather than from a general inability to generalize across languages.
This observation is consistent with recent findings that synthetic speech attribution models may accidentally encode speaker-related information, entangling speaker characteristics with synthesis source representations~\cite{xuan2026disentangling}.

We now examine the remaining cross-lingual gap under the Filtered setting.
\Cref{tab:_frequent_lang} reports per-language results of ProtoRN18 L4P6, the best-performing model in the previous evaluation, for the ten languages with at least five covered architecture classes.
Most languages show a moderate accuracy drop relative to the English-matched subset.
Russian is a notable outlier, partly explained by a dramatic absence of true positive samples for Meta's \texttt{MMS-TTS} generator in the corresponding evaluation subset.

\subsection{Open-set Performance}
\label{subsec:res_4}

\begin{table}
\caption{Open-set performance. Results are reported in terms of AUC, FPR at 95\% TPR and Unknown Rejection Rate (URR). URR threshold computed on the validation set.}
\label{tab:open_set}
\resizebox{\columnwidth}{!}{
\begin{tabular}{lccc}
\hline
\toprule
& \textbf{Avg AUC $\uparrow$} & \textbf{Avg FPR@TPR95 $\downarrow$} & \textbf{Avg URR $\uparrow$} \\
\midrule
ProtoRN18 L4P3 & \textbf{89.84\%} & \textbf{29.60\%} & \textbf{68.83\%} \\
ProtoRN18 L4P6 & 83.76\% & 32.43\% & 67.16\% \\
\bottomrule
\end{tabular}}
\end{table}

\begin{table*}[t]
\caption{Unknown class detection results and top-3 predicted source models by ProtoRN18 L4P3.}
\label{tab:unknown_classes}
\resizebox{\textwidth}{!}{
\begin{tabular}{lccclll}
\hline
\toprule
& 
\textbf{AUC $\uparrow$} &
\textbf{FPR@TPR95 $\downarrow$} &
\textbf{URR $\uparrow$} &
\textbf{Top 1} &
\textbf{Top 2} &
\textbf{Top 3} \\
\midrule
Index-TTS           & 93.58\% & 18.80\% & 79.60\% & griffin\_lim (33.05\%) & FireRedTTS (20.15\%) & Llasa (9.05\%) \\
Kitten-TTS          & 88.36\% & 34.15\% & 64.95\% & Spark-TTS-0.5B (43.00\%) & Llasa (42.10\%) & Chatterbox (12.95\%) \\
MiniCPM-o-2.6       & 98.45\% & 4.70\%  & 94.20\% & FireRedTTS (13.10\%) & orpheus-tts-0.1-finetune (11.50\%) & MegaTTS3 (9.10\%) \\
Kyutai-TTS          & 95.47\% & 17.10\% & 80.90\% & sesame\_csm (34.90\%) & parler\_tts (28.90\%) & kokoro (6.30\%) \\
Microsoft VibeVoice & 93.22\% & 27.35\% & 70.00\% & Spark-TTS-0.5B (30.30\%) & Mars5 (27.65\%) & MegaTTS3 (21.80\%) \\
Veena               & 56.29\% & 97.30\% & 2.40\%  & orpheus-tts-0.1-finetune (99.70\%) & Chatterbox (0.10\%) & f5-tts (0.10\%) \\
Higgs-Audio-V2      & 92.37\% & 22.50\% & 76.50\% & Llasa (53.00\%) & Spark-TTS-0.5B (26.50\%) & griffin\_lim (6.40\%) \\
VoxCPM              & 92.92\% & 28.10\% & 70.10\% & Spark-TTS-0.5B (49.80\%) & Llasa (20.90\%) & griffin\_lim (10.10\%) \\
OuteTTS             & 91.61\% & 25.10\% & 73.70\% & Llasa (50.60\%) & Spark-TTS-0.5B (19.70\%) & Mars5 (15.10\%) \\
Qwen2.5-Omni        & 95.97\% & 13.50\% & 85.60\% & Llasa (28.10\%) & Resemble.ai (27.40\%) & MegaTTS3 (24.90\%) \\
Voxtream            & 89.95\% & 37.00\% & 59.20\% & sesame\_csm (76.80\%) & FireRedTTS (7.60\%) & f5-tts (3.90\%) \\
\bottomrule
\end{tabular}}
\end{table*}

In this experiment, we evaluate the ability of our models to reject samples from previously unseen classes in a realistic open-set scenario. 
We consider \num{11} unknown \gls{tts} families introduced between MLAAD v7 and v8 (see \Cref{subsec:dataset}).

Rejection requires a score computable without access to the ground-truth label, since a genuinely unknown sample has no true class to measure similarity against.
We considered two such scores derived from the model's prototype similarities: the \textit{maximum} similarity across all prototypes regardless of class, and the \textit{margin} between the highest and second-highest per-class maximum similarity.
The latter substantially outperformed the former in preliminary experiments.
This is consistent with the intuition that a genuinely unknown sample tends to match multiple known classes weakly rather than any single class strongly, a property the margin captures directly while raw similarity does not.
We therefore adopt this margin as the rejection score for all results reported in this section.

\Cref{tab:open_set} compares open-set rejection performance between ProtoRN18 L4P3 and ProtoRN18 L4P6. 
In addition to the \gls{auc} and the derived FPR@TPR95, we report the Unknown Rejection Rate (URR), computed using a rejection threshold selected on the validation set to provide a more deployment-oriented evaluation.
Unlike the closed-set setting, L4P3 consistently outperforms L4P6 across all metrics. 
This suggests that, while additional prototypes improve closed-set discrimination, they also broaden the learned feature space, increasing the probability that an unseen generator matches a known prototype closely enough to be accepted.
We therefore use L4P3 for the next analysis.

\Cref{tab:unknown_classes} reports per-generator results for each of the unknown architectures. 
Rejection performance varies substantially: MiniCPM-o-2.6 is rejected almost perfectly (URR \num{94.20}\%), whereas Veena is rarely rejected (URR \num{2.40}\%) despite a moderate AUC (\num{56.29}\%). 
A deeper analysis revealed that, for Veena, \num{99.70}\% of samples are attributed to a single known class, \texttt{orpheus-tts-0.1-finetune}.
This consistent collapse onto a single known class suggests genuine similarity between the two systems, which are indeed both built upon a Llama 3B backbone.
More broadly, top-1 attributions in \Cref{tab:unknown_classes} are dominated by a small set of known classes, i.e., \texttt{Llasa}, \texttt{Spark-TTS-0.5B}, and \texttt{sesame\_csm}. 
This pattern is again consistent with shared synthesis paradigms between known and held-out systems, suggesting that open-set failures in prototypical networks may indeed be interpretable to some extent. 

\section{Conclusions}
\label{sec:conclusions}
In this work, we investigated prototypical networks as an interpretable alternative for synthetic speech attribution. 
Our results show that prototype-based reasoning does not necessarily require sacrificing performance.
We further evaluated the framework under cross-lingual and open-set conditions. 
We showed that part of the cross-lingual degradation was linked to limited speaker coverage in the training data, highlighting the importance of speaker diversity for robust source representations. 
In the open-set scenario, prototype similarities enabled effective rejection of several unseen generators while exposing interpretable failure cases.
Overall, prototype-based models can maintain strong attribution accuracy while providing insights into the evidence supporting their decisions.

\bibliographystyle{IEEEtran}
\bibliography{bibliography.bib}

\end{document}